\documentclass[10pt,twocolumn,twoside]{elsart}

\usepackage{graphicx}
\usepackage{amsmath}

\begin{document}

\begin{frontmatter}

\title{Azimuthal asymmetry of direct photons in intermediate energy heavy-ion collisions}

\author{G. H. Liu$^{a,b}$,} \author{ Y. G. Ma$^{a}$, } \ead{
ygma@sinap.ac.cn. Corresponding author.  }
\author{  X. Z. Cai$^{a}$,  D. Q. Fang$^{a}$,  W. Q.
Shen$^{a}$, W. D. Tian$^{a}$, K. Wang$^{a}$}
\address{$^a$ Shanghai Institute of Applied Physics, Chinese Academy of
Sciences, Shanghai 201800, China\\
$^b$ Graduate School of the Chinese Academy of Sciences, Beijing
100080, China}

\date{\today}

\begin{abstract}

Hard photon emitted from  energetic heavy ion collisions is of
very interesting since it does not experience the late-stage
nuclear interaction, therefore it is useful to explore the
early-stage information of matter phase. In this work,  we have
presented a first calculation of azimuthal asymmetry,
characterized by directed transverse flow parameter $F$ and
elliptic asymmetry coefficient $v_2$, for proton-neutron
bremsstrahlung hard photons in intermediate energy heavy-ion
collisions. The positive $F$ and negative $v_2$ of direct photons
are illustrated and they seem to be anti-correlated to the
corresponding free proton's flow.

\end{abstract}
\end{frontmatter}

\maketitle

The properties of nuclear matter at different temperatures or
densities, especially the derivation of the Equation-of-State
(EOS) of nuclear matter, are one of the foremost challenges of
modern heavy-ion physics. Since heavy ion collisions provide up to
now the unique means to form and investigate hot and dense nuclear
matter in the laboratory, many experimental and theoretical
efforts are under way towards this direction. Because of their
relatively high emission rates, nucleons, mesons, light ions and
intermediate mass fragments, produced and emitted in the
reactions, are conveniently used to obtain information on the
reaction dynamics of energetic heavy ion collisions. However,
these probes interact strongly with the nuclear medium such that
the information they convey may bring a blurred image of their
source. Fortunately, energetic photons offer an attractive
alternative to the hadronic probes \cite{schutz}. Photons
interacting only weakly through the electromagnetic force with the
nuclear medium are not subjected to distortions by the final state
(neither Coulomb nor strong) interactions. They therefore deliver
an undistorted picture of the emitting source. For hard photons,
defined as $\gamma$-rays with energies above $30 MeV$ in this
paper, many experimental facts supported by model calculations
\cite{schutz,cassing,nifenecker} indicate that in intermediate
energy heavy-ion collisions they are mainly emitted during the
first instants of the reaction in incoherent proton-neutron
bremsstrahlung collisions, $p+n \rightarrow p+n+\gamma$, occurring
within the participant zone. This part of hard photons are called
as direct photon. Direct hard photons have thus been exploited to
probe the pre-equilibrium conditions prevailing in the initial
high-density phase of the reaction \cite{wada,schmidt}. Aside from
the dominant production of hard photons in first-chance p-n
collisions, a significant hard-photon production in a later stage
of heavy-ion reactions, called as thermal photons, are also
predicted by the \textit{Boltzmann-Uehling-Uhlenbeck}
(\textit{BUU}) theory \cite{schutz1,schutz2}. These thermal
photons are emitted from a nearly thermalized source and still
originate from bremsstrahlung production by individual p-n
collisions, which was also confirmed by the experiments at last
decade \cite{martinez,enterria}.

In this work, we take the \textit{BUU} transport model improved by
Bauer \cite{Bauer}. The isospin dependence was incorporated into
the model through the initialization and the nuclear mean field.
The nuclear mean field $U$ including isospin symmetry terms is
parameterized as
\begin{equation}
  U(\rho,\tau_{z}) = a(\frac{\rho}{\rho_{0}}) +
  b(\frac{\rho}{\rho_{0}})^{\sigma} + C_{sym} \frac{(\rho_{n} -
    \rho_{p})}{\rho_{0}}\tau_{z},
\end{equation}
where $\rho_0$ is the normal nuclear matter density; $\rho$,
$\rho_n$, and $\rho_p$ are the nucleon, neutron and proton
densities, respectively; $\tau_z$ equals 1 or -1 for neutrons and
protons, respectively; The coefficients $a$, $b$ and $\sigma$ are
parameters for nuclear equation of state. $C_{sym}$ is the
symmetry energy strength due to the density difference of neutrons
and protons in nuclear medium, which is important for asymmetry
nuclear matter (here $C_{sym} = 32$ $MeV$ is used), but it is
trivial for the symmetric system studied in the present work.

For the calculation of the elementary double-differential hard
photon production cross sections on the basis of individual
proton-neutron bremsstrahlung, the hard-sphere collision was
adopted  from \textit{Ref.} \cite{Jackson}, and modified as in
\textit{Ref.} \cite{Cassing2} to allow for energy conservation.
The double differential probability is given by
\begin{equation}
\frac{d^2\sigma^{elem}}{dE_{\gamma}d\Omega_{\gamma}}=\alpha\frac{R^2}{12\pi}\frac{1}{E_{\gamma}}(2\beta_f^2+3sin^2\theta_{\gamma}\beta_i^2).
\label{ddcs}
\end{equation}
Here $R$ is the radius of the sphere, $\alpha$ is the fine
structure constant, $\beta_i$ and $\beta_f$ are the initial and
final velocity of the proton in the proton-neutron center of mass
system, and $\theta_{\gamma}$ is the  angle between incident
proton direction and photon emitting direction. More details for
the model can be found in Ref.~\cite{Bauer}.

\begin{figure}
  \includegraphics[width=0.4\textwidth]{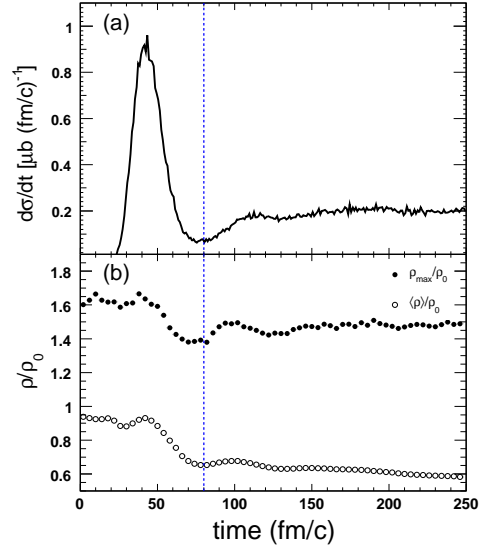}
  \vspace{0.1truein}
  \caption{\footnotesize (a) Time evolution of hard photon production rate
  for the reaction  $^{40}Ca + ^{40}Ca$ collisions at 30 $MeV/nucleon$ for semi-central events (40--60\%).
    (b) Time evolution of reduced
    maximum density $\rho_{max}/\rho_0$ (closed circles) and reduced average
    density $\langle \rho \rangle / \rho_0$ (open circles) of the whole reaction system in the same reaction.
    The blue dashed line represents the time when the system ends up till the first expansion stage, and in the panel (a) it separates
    direct photons (on the left side) and thermal photons (on the right side).
  }\label{time}
\end{figure}

In this paper, we simulate the reaction of $^{40}Ca + ^{40}Ca$
collisions at $30 MeV/nucleon$, and use the $EOS$ with the
compressibility $K$ of 235 $MeV$ ($a$ = -218 $MeV$, $b$ = 164
$MeV$, $\sigma$ = 4/3) for the nuclear mean field $U$. As a first
attempt to extract the photon's azimuthal asymmetry, we only take
the semi-central events $(40-60\%)$ as an example in this Letter.

In Fig.~\ref{time} we show the time evolution of production rate
of bremsstrahlung hard photons as well as the time evolution of
system densities, including both maximum (closed circles) and
average density (open circles). We found that hard-photon
production is sensitive to the density oscillations of both the
maximum and the average density during the whole reaction
evolution. When the density of collision system increases, that is
in the compression stage, the system produces more hard photons.
In contrary, when the system expends, the hard photon production
decreases. Actually, the density oscillations of the colliding
heavy ions systems can be observed in the experiments via
hard-photon interferometry measurements \cite{marques,schutz1}.
Apparently, hard photons are mostly produced at the early stage of
the reaction. Combining the time evolution of the nuclear density,
we know that this part of hard photons are dominantly emitted from
the stage of the first compression and expansion of the system.
Thereafter we call these photons, emitted before the time of the
first maximum expansion of the system ($t = 80$ $fm/c$ in this
reaction), as direct photons (on the left side of blue dashed line
in Fig.~\ref{time}(a)). It is also coincident with the definition
of direct photons above. And we call the residual hard photons
produced in the later stage as thermal photons (on the right side
of blue dashed line in Fig.~\ref{time}(a)). So in the simulation,
we can identify the produced photon as direct or thermal photon by
the emitting time. Because of the sensitivity to the density
oscillations of colliding system, hard photon may be sensitive to
the nuclear incompressibility \cite{schutz1,schutz2}.

It is well known that collective flow is an important observable
in heavy ion collisions and it can bring some essential
information of the nuclear matter, such as the nuclear equation of
state
\cite{Olli,Voloshin,Sorge,Danile,Ma,Zheng,Gale,INDRA,Yan,Chen}.
Anisotropic flow is defined as the different $n-th$ harmonic
coefficient $v_n$ of the Fourier expansion for the particle
invariant azimuthal distribution \cite{Voloshin}:
\begin{equation}
  \frac{dN}{d\phi}\propto{1+2\sum^\infty_{n=1}{v_n\cos(n\phi)}},
  \label{Fourier}
\end{equation}
where $\phi$ is the azimuthal angle between the transverse
momentum of the particle and the reaction plane. Note that the
$z$-axis is defined as the direction along the beam and the impact
parameter axis is labelled as $x$-axis. Anisotropic flows
generally depend on both particle transverse momentum and
rapidity, and for a given rapidity the anisotropic flows at
transverse momentum $p_t$ ($p_t = \sqrt{p_x^2+p_y^2}$) can be
evaluated according to
\begin{equation}
  v_{n}(p_t) = \langle cos(n\phi) \rangle,
\end{equation}
where $\langle \cdots \rangle$ denotes average over the azimuthal
distribution of particles with transverse momentum $p_t$, $p_x$
and $p_y$ are projections of particle transverse momentum in and
perpendicular to the reaction plane, respectively. The first
harmonic coefficient $v_1$ is called directed flow parameter. The
second harmonic coefficient $v_2$ is called the elliptic flow
parameter $v_2$, which measures the eccentricity of the particle
distribution in the momentum space.

In relativistic heavy-ion collisions azimuthal asymmetry of hard
photons have been recently reported in the experiments and
theoretical calculations \cite{Aggarwal,Adams,Turbide,Heinz}. It
shows a very useful tool to explore the properties of hot dense
matter. However, so far there is still neither experimental data
nor theoretical prediction on the azimuthal asymmetry of hard
photons in intermediate energy heavy ion collisions. Does the
direct photon also exist azimuthal asymmetry so that it leads to
non-zero directed transverse flow  or elliptic asymmetry
parameters in the intermediate energy range? Moreover we know that
direct photons mostly originate from bremsstrahlung produced in
individual proton-neutron collisions, and free nucleons are also
emitted from nucleon-nucleon collisions. Does the azimuthal
asymmetry of the direct photons  correlate with the one of free
nucleons? To answer the above question, we focus on the azimuthal
asymmetry analysis for both photons and protons in this Letter.

\begin{figure}
  \vspace{-0.1truein}
  \includegraphics[width=0.40\textwidth]{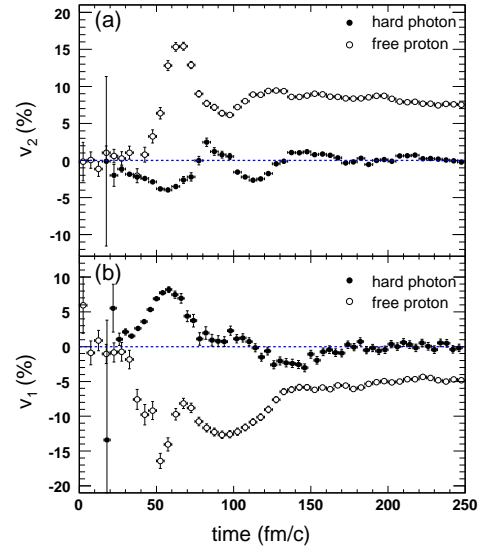}
   \vspace{0.1truein}
  \caption{\footnotesize The time evolution of $v_2$ (a) and $v_1$ (b) for
    hard photons (closed circles) and free protons (open circles).
  }\label{flow_time}
\end{figure}

Fig.~\ref{flow_time} shows the time evolution of the directed flow
parameter $v_1$ and elliptic flow parameter $v_2$ for hard photons
and free protons. Before we take further calculation and
explanation, people should be cautious about the word of "flow"
for photons. Since flow is associated with collectivity caused by
multiple interactions, which are exhibited by the nucleons, but
not by the photons. The photon emission pattern is basically a
result of the nucleon flow, and not a photon flow per se. However,
However, in order to compare the results between photons and
protons, we still called $v_1$ as directed flow parameter and
$v_2$ elliptic flow parameters for photons somewhere in texts.
Considering the nearly symmetric behavior for directed flow
parameter ($v_1$) versus rapidity, here we calculate the average
$v_1$ over only the positive rapidity range, which can be taken as
a measure of the directed transverse flow parameter. For emitted
protons (open circles in the figure) which are experimental
measurable and are identified in our BUU calculation as those with
local densities less than $\rho_0/8$, the onset of flows occurs
around $t$ = 30 $fm/c$ before that the system is mostly in fusion
stage and protons are seldom emitted. The negative directed flow
parameter $v_1$ of free protons essentially stems from the
attractive mean field. Up till $ t \sim 120 fm/c$ when the system
is in the freeze-out stage, the directed flows become saturate.
For the elliptic asymmetry parameter $v_2$ of free protons, the
positive values indicate of the preferential in-plane emission
driven by the rotational collective motion due to the attractive
mean filed. Similarly, the elliptic asymmetry parameter becomes
saturate in the freeze-out stage. However, there are obvious
difference for proton-neutron bremsstrahlung photons (solid
circles in the figure) in comparison to protons. Contrary to the
negative directed transverse flow and positive flow, directed
photons shows the positive $v_1$ and the negative $v_2$ before $t
= 80fm/c$, i.e. the azimuthal anisotropy is shifted by a phase of
$\pi$/2. The times corresponding to the peak or valley values of
flows roughly keep synchronized with the compression or expansion
oscillation of the system evolution. For the late-stage thermal
photons after $t = 170 fm/c$ the azimuthal asymmetry vanishes,
i.e. $v_1$ and $v_2$ fades-out.

From the above calculations, we learn that thermal photons in the
later stage of reaction are emitted from a more thermalized
system, they prefer more isotropic emission (i.e. the vanishing
"flow" parameters) than direct ones produced in the
pre-equilibrium stage. Thereafter we only consider direct photons
to discuss the azimuthal asymmetry results. For protons, we take
the values of flows when the system has been already in the
freeze-out time at 180 $fm/c$.

\begin{figure}
  \vspace{-0.1truein}
\includegraphics[width=0.48\textwidth]{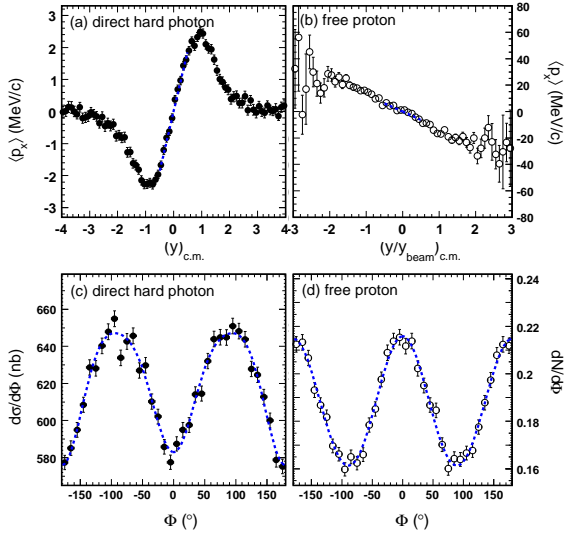}
  \vspace{0.1truein}
  \caption{\footnotesize (a) Average in-plane transverse momentum of direct photons as a function of $c.m.$ rapidity
for semi-central events $(40-60\%)$.
   The dashed line segment is a fit over the mid-rapidity region $-0.5 \leq y_{c.m.}\leq 0.5$.
   (b) Same as the panel (a) but for free protons.
   The dashed line segment is a fit over the mid-rapidity region $-0.5 \leq(y/y_{beam})_{c.m.}\leq 0.5$.
   (c) and (d) are the azimuthal distributions of direct photons and free protons, respectively, and both of them are
   fitted to the $4th$ order Fourier expansion.
  }\label{flow}
  %\label{direct_flow}
\end{figure}

The directed transverse flow parameter at mid-rapidity can be also
defined by the slope:
  $F = \left. \frac{d \langle p_x \rangle}{d(y)_{c.m.}}
  \right|_{(y)_{c.m.}=0}$,
where $(y)_{c.m.}$ is the rapidity of particles in the center of
mass and $\langle p_x \rangle$ is the mean in-plane transverse
momentum of photons or protons in a given rapidity region. In
Fig.~\ref{flow}(a) and (b), we show  $\langle p_x \rangle$ plotted
versus the \textit{c.m.} rapidity $y_{c.m.}$ for direct photons
(a) as well as $\langle p_x \rangle$ plotted versus the reduced
\textit{c.m.} rapidity $(y/y_{beam})_{c.m.}$ for free protons (b).
The errors shown are only statistical. A good linearity was seen
in the mid-rapidity region $(-0.5, 0.5)$ and the slope of a linear
fit can be  defined as the directed transverse flow parameter. The
extracted value of the directed transverse flow of direct photons
is about $+3.7$ $MeV/c$, and that of free protons is about $-12.4$
$MeV/c$. Thus direct photons do exist the directed transverse
asymmetry even though the absolute value is smaller than the
proton's flow, and its sign is just opposite to that of free
protons.

\begin{figure}
\includegraphics[width=0.5\textwidth]{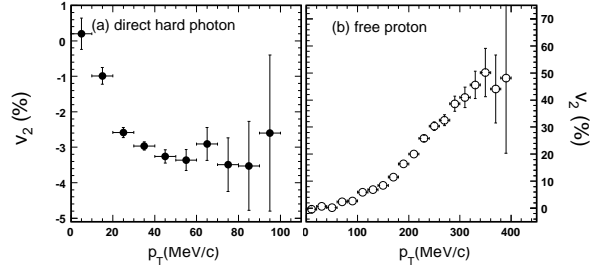}
  \vspace{0.1truein}
  \caption{\footnotesize $v_2$  as a function of transverse momentum ($p_{T}$) for direct photons (a) and free protons (b).
  }\label{v2Pt}
\end{figure}

As Eq.~\ref{Fourier} shows, elliptic flow is defined as the second
harmonic coefficient $v_2$ of an azimuthal Fourier expansion of
the particle invariant distribution. In order to extract the value
of elliptic asymmetry coefficient $v_2$ and reduce the error of
fits, we fit the azimuthal distribution to the $4th$ order Fourier
expansion. Shown in Fig.~\ref{flow}(c) and (d), direct photons
demonstrate out-of-plane enhancement  and the $v_{2}$ is about
$-2.7\%$. Whereas, for free protons, azimuthal distribution
displays the preferential in-plane emission and the $v_2$ is about
$+7.2\%$. Furthermore, we can extract the transverse momentum
dependence of the elliptic asymmetry coefficient $v_2$.
Fig.~\ref{v2Pt} shows $v_2$ of direct photons (a) and free protons
(b) as a function of transverse momentum $p_{T}$. Similar to the
directed transverse flow parameter, the values of elliptic
asymmetry coefficient $v_2$ of direct photons and free protons
also have the opposite signs at this reaction energy, i.e.
reflecting a different preferential transverse emission in the
direction of out-of-plane or in-plane, respectively. Meanwhile,
the absolute values of $v_2$ for photons are smaller than the
proton's values as the behavior of transverse flow. Except the
opposite sign, we see that both $v_2$ have similar tendency with
the increase of $p_{T}$, i.e., their absolute values increase at
lower $p_{T}$, and become gradually saturated, especially for
direct photons.

To explain the above anti-correlation of anisotropic emission
between direct photons  and free protons, we should note that
direct photons originate from the individual proton-neutron
collisions. As Eq.~\ref{ddcs} shows, we can roughly consider that
in the individual proton-neutron center of mass system, in
directions perpendicular to incident proton velocity, i.e.
$\theta_{\gamma} = \pi/2$, the probability of hard photon
production is much larger than that in the parallel direction,
i.e. $\theta_{\gamma} = 0$, which is in agreement with the
theoretical calculations and the experiments
\cite{Herrmann,Safkan}, that causes hard photon preferential
emission perpendicular to the motion plane of corresponding
nucleons. As a whole, the azimuthal anisotropy of hard photons is
shifted by a phase of $\pi/2$ with respect to that associated with
the anisotropy of nucleons, leading to the opposite signs of the
values of $F$ and $v_2$ between them. Consequently, azimuthal
anisotropic emission of hard photon and free nucleon  are
anti-correlated, presenting the opposite behavior.

In conclusion, we have presented a first calculation of azimuthal
asymmetry, both directed and elliptic asymmetry, for direct
photons produced by proton-neutron bremsstrahlung from
intermediate energy heavy-ion collisions. It was, for the first
time, presented that in the intermediate energy heavy-ion
collisions the proton-neutron bremsstrahlung hard photon shows
non-zero directed transverse flow parameter and elliptic asymmetry
coefficient which have opposite sign to the corresponding free
proton flow parameters. The time evolutions of azimuthal
parameters $v_1$ and $v_2$ of hard photons exhibit rich structures
as the density oscillation of the system during the
pre-equilibrium and thermalization stage of reaction system.
Therefore direct photons can server for a good probe to nuclear
matter properties. Considering that hard photons are dominantly
produced by individual neutron-proton bremsstrahlung, so they are
sensitive to the in-medium neutron-proton cross section, but not
to the in-medium proton-proton or neutron-neutron cross section,
that can be advantaged in the isospin dependent study of in-medium
nucleon-nucleon cross section by direct photons. Of course,
systematic studies of the influences from equation of state,
in-medium nucleon-nucleon cross section, impact parameter and
incident energy etc on the azimuthal asymmetry of direct photon
should be carried out. The progress along this line is underway.

This work was partially supported by the National Basic Research
Program of China (973 Program) under Contract No. 2007CB815004,
Shanghai Development Foundation from Science and Technology under
Grant Numbers 06JC14082 and 06QA14062, the National Natural
Science Foundation of China under Grant No 10535010 and 10775167.

%\footnotesize
{}

\end{document}